\documentstyle[12pt,epsfig]{article}                                                    
                                                    
\newlength{\dinwidth}                                                    
\newlength{\dinmargin}                                                    
\def\lapproxeq{\lower .7ex\hbox{$\;\stackrel{\textstyle                                                    
<}{\sim}\;$}}                                                    
\def\gapproxeq{\lower .7ex\hbox{$\;\stackrel{\textstyle                                                    
>}{\sim}\;$}}                    
\newcommand{\porpbar}
{\!\,^{\scriptscriptstyle(}$\mbox{$\bar{p}$}$\,^{\scriptscriptstyle)}}                                
\def\be{\begin{equation}}                                                    
\def\ee{\end{equation}}                                                    
\def\bea{\begin{eqnarray}}                                                    
\def\eea{\end{eqnarray}}

\begin{document}                                                    
\titlepage                                                    
\begin{flushright}                                                    
IPPP/00/10 \\
DCTP/00/74 \\                                                    
20 April 2001 \\                                                    
\end{flushright}                                                    
                                                    
\vspace*{2cm}                                                    
                                                    
\begin{center}                                                    
{\Large \bf Double-diffractive processes in high-resolution} \\           
           
\vspace*{0.5cm}           
{\Large \bf missing-mass experiments at the Tevatron}                                                    
                                                    
\vspace*{1cm}                                                    
V.A. Khoze$^a$, A.D. Martin$^a$ and M.G. Ryskin$^{a,b}$ \\                                                    
                                                   
\vspace*{0.5cm}                                                    
$^a$ Department of Physics, University of Durham, Durham, DH1 3LE \\                                                   
$^b$ Petersburg Nuclear Physics Institute, Gatchina, St.~Petersburg, 188300, Russia            
\end{center}                                                    
                                                    
\vspace*{2cm}                                                    
                                                    
\begin{abstract}                                                    
We evaluate, in a model-independent way, the signal-to-background ratio for 
Higgs$\rightarrow b\bar{b}$ detection in exclusive double-diffractive events at the Tevatron 
and the LHC.  For the missing-mass approach to be able to identify the Higgs boson, it will be 
necessary to use a central jet detector and to tag $b$ quark jets.  The signal is predicted to be 
very small at the Tevatron, but observable at the LHC.  However we note that the 
background, that is double-diffractive dijet production, may serve as a unique gluon factory.  
We also give estimates for the double-diffractive production of $\chi_c$ and $\chi_b$ mesons 
at the Tevatron.  We emphasize that a high-resolution missing-mass measurement, on its own, 
is insufficient to identify rare processes.
\end{abstract}                                          
     
\newpage           
\section{Introduction}         
>From several points of view it looks appealing to study processes with two large rapidity gaps 
in high energy hadron collisions.  Applications of such processes, generated, for example, by 
\lq Pomeron-Pomeron\rq\ collisions, embrace both searches for New Physics (such as the 
Higgs boson) and dedicated analyses of conventional physics, including the investigation of 
subtle aspects of QCD.  The attractiveness of the approach is motivated by the spectacularly 
clean experimental signatures and the possibility to clearly differentiate between different 
production mechanisms.

Events with large rapidity gaps may be selected either by using a calorimeter or by detecting 
leading protons with beam momentum fractions $x$ close to 1.  If the momenta of the leading 
protons can be measured with very high accuracy then a particle (or system) produced by the 
double-diffractive mechanism may be observed as a peak in the spectrum of the missing-mass 
($M$) distribution.  Indeed, it has recently been proposed to search for the Higgs 
boson\footnote{The possibility of a high resolution missing-mass search for the Higgs boson 
in exclusive double rapidity gap events was considered in Ref.~\cite{JE}.} by 
measuring the outgoing fast proton and antiproton in Run II of the Tevatron with 
extremely good precision corresponding to a missing-mass resolution $\Delta M \simeq 
250~{\rm MeV}$ \cite{AR}.  To ascertain whether the sought after Higgs signal can be seen, 
it is crucial to evaluate the background.  Recall that the inclusive search for an intermediate 
mass Higgs, that is $pp$ or $p\bar{p} \rightarrow HX$ with $H \rightarrow b\bar{b}$, has an 
extremely small signal-to-background ratio, which makes this process impossible to observe.

In Section~2 we briefly recall the QCD mechanism for the double-diffractive production of a 
system of large invariant mass $M$.  We use this formalism in Section~3 to study the 
background for double-diffractive $H \rightarrow b\bar{b}$ production, and to show that the 
signal-to-background ratio can be estimated in a practically model-independent way.  Then in 
Section~4 we discuss 
double-diffractive production of $\chi_b$ (and $\chi_c$) mesons.  
We present their expected cross sections at the Tevatron.  In Section~5 we discuss 
another attractive possibility.  That is, to use the double-diffractive production of a dijet 
system as a \lq\lq gluon factory\rq\rq, which generates huge numbers of essentially pure gluon 
jets in a clean environment.  In Section~6 we emphasize that a high-resolution missing-mass 
measurement on its own may not yield a sharp peak for a rare process, if care is not taken to 
account for QED radiation.  Finally, in Section~7, we present our conclusions.

\section{The mechanism for double-diffractive production}

We wish to estimate the cross section for high energy reactions of the type
\be
\label{eq:a1}
pp \; \rightarrow \; p \: + \: M \: + \: p,
\ee
and similarly for $p\bar{p}$, where the \lq plus\rq\ signs indicate the presence of large 
rapidity gaps.  To be precise, we calculate the rate for the double-diffractive exclusive 
production of a system of large invariant mass $M$, for example, a Higgs boson.  In all 
models \cite{KMR2}--\cite{KMR3} 
the amplitude for the double-diffractive process is described by Fig.~1, where the hard 
subprocess $gg \rightarrow M$ is initiated by gluon-gluon fusion and the second $t$-channel 
gluon is needed to screen the colour flow across the rapidity gap intervals.  In other words the 
Pomeron is modelled by two-gluon exchange.

One major difference between the various theoretical approaches concerns the specification 
of the exchange gluons.  {\it Either} non-perturbative gluons are used in which the propagator 
is modified so as to reproduce the total cross section \cite{HIGGS,EML}, {\it or} a 
perturbative QCD estimate is made \cite{KMR3} using an unintegrated, skewed gluon 
density that is determined from conventional gluons obtained in global parton analyses.  
However it has been emphasized \cite{BE} (see also \cite{KMR1,KMR3}) that the non-
perturbative normalisation based on the value of the elastic or total cross section fixes the 
diagonal gluon density at $\hat{x} \sim \ell_T/\sqrt{s}$ where the transverse momentum 
$\ell_T$ is small, namely $\ell_T < 1~{\rm GeV}$.  Thus the value of $\hat{x}$ is even 
smaller than
\be
\label{eq:a2}
x^\prime \; \approx \; \frac{Q_T}{\sqrt{s}} \; \ll \; x \; \approx \; \frac{M}{\sqrt{s}},
\ee
where the variables are defined in Fig.~1.  However, the gluon density grows as $x 
\rightarrow 0$ and so the use of a non-perturbative normalisation will lead to an 
overestimation of double-diffractive cross sections.

Of course the fusion of the two energetic gluons into the high mass state in Fig.~1 is generally 
accompanied by the emission of soft gluons which may populate the rapidity gaps.  The basic 
mechanism to suppress this effect is shown in Fig.~1, where the second $t$-channel gluon, 
which screens the colour, does not couple to the produced state of mass $M$ and has typical 
values of $Q_T$ which are much smaller than $M$ but yet are large enough (for sufficiently 
large $M$) to screen soft gluon emission and to justify the applicability of perturbative QCD.

Recently the $p\porpbar \rightarrow p + H + \porpbar$ cross section has been calculated to 
single $\log$ accuracy \cite{KMR3}.  The amplitude is
\be
\label{eq:b2}
{\cal M} \; = \; A \pi^3 \: \int \: \frac{d^2 Q_T}{Q_T^4} \: f_g (x_1, x_1^\prime, Q_T^2, 
M_H^2/4) \: f_g (x_2, x_2^\prime, Q_T^2, M_H^2/4), 
\ee
where the $gg \rightarrow H$ vertex factor $A^2$ is given by (\ref{eq:a4}) below, and the 
unintegrated gluon densities are related to the conventional distributions by
\be
\label{eq:c2}
f_g (x, x^\prime, Q_T^2, M_H^2/4) \; = \; R_g \: \frac{\partial}{\partial \ln Q_T^2} \left [ 
\sqrt{T (Q_T, M_H/2)} \: xg (x, Q_T^2) \right ].
\ee
The factor $R_g$ is the ratio of the skewed $x^\prime \ll x$ integrated gluon distribution to 
the conventional one.  $R_g \simeq 1.2 (1.4)$ at LHC (Tevatron) energies.  The 
bremsstrahlung survival probability $T^2$ is given by
\be
\label{eq:d2}
T (Q_T, \mu) \; = \; \exp \left (- \: \int_{Q_T^2}^{\mu^2} \: \frac{dk_T^2}{k_T^2} \: 
\frac{\alpha_S (k_T^2)}{2 \pi} \: \int_0^{1 - k_T/\mu} \: dz \: \left [z \: P_{gg}(z) \: + \: 
\sum_q \: P_{qg} (z) \right ] \right ),
\ee
and strongly suppresses the infrared contribution to the $Q_T$ integration of (\ref{eq:b2}).  
The factor $\sqrt{T}$ arises in (\ref{eq:c2}) because the survival probability is only relevant 
to the hard gluon exchanges in Fig.~1.  In addition to this suppression due to the probability 
\lq\lq $T^2$\rq\rq\ that the $pp \rightarrow p + H + p$ rapidity gaps survive population by 
extra gluons from the hard process, we must also include the probability $S^2$ that the gaps 
are not filled by secondaries produced in soft rescattering between the protons, that is by an 
underlying interaction.  We estimate the $p\bar{p} \rightarrow p + H + \bar{p}$ event rate at 
the Tevatron in Section~3.3.

\section{Dijet background to double-diffractive Higgs production}

To use the \lq missing-mass\rq\ method to search for an intermediate mass Higgs boson, via 
the $H \rightarrow b\bar{b}$ decay mode, we have to estimate the QCD background which 
arises from the production of a pair of jets with invariant mass about $M_H$.  If we assume 
that the Higgs boson is produced by the $gg \rightarrow H$ fusion mechanism then the 
signal-to-background ratio is just given by the ratio of the appropriate matrix elements 
squared for the $gg \rightarrow H$ and $gg \rightarrow$~dijet subprocesses.

\subsection{Gluon dijet background}

We begin by considering the double-diffractive colour-singlet production of a pair of high 
$E_T$ gluons with rapidities $\eta_1$ and $\eta_2$.  The $gg \rightarrow gg$ subprocess 
cross section is \cite{KMR2,BC}
\bea
\label{eq:a3}
\frac{d \hat{\sigma}}{d^2 p_T} & = & \frac{9 \alpha_S^2}{4 p_T^4} \: \frac{1}{2 
p_T^2 \: \sinh \: \Delta \eta} \: \frac{dM^2}{d (\Delta \eta)} \nonumber \\
& & \\
& = & \frac{9 \alpha_S^2}{8 p_T^6} \: \left ( \frac{M^4}{4 p_T^4} \: - \: 
\frac{M^2}{p_T^2} \right )^{- \frac{1}{2}} \: \frac{dM^2}{d (\Delta \eta)}, \nonumber
\eea
where $M$ is the invariant mass of the dijet system, $p_T$ is the transverse momentum of 
the jets, and $\Delta \eta = |\eta_1 - \eta_2|$ is the jet rapidity difference.  Note that, since the 
outgoing proton and antiproton are at small angles relevant to the respective beams, 
$\mbox{\boldmath $p$}_{1T} = -\mbox{\boldmath $p$}_{2T}$.

The background (\ref{eq:a3}) should be compared to the double-diffractive $gg \rightarrow 
H$ signal
\be
\label{eq:a4}
\frac{A^2}{4} \; = \; \frac{\sqrt{2}}{4} \: G_F \: \alpha_S^2 \: \frac{N}{9 \pi^2} \; \simeq \; 
\frac{\sqrt{2}}{36 \pi^2} \: G_F \: \alpha_S^2,
\ee
where $G_F$ is the Fermi coupling and $N \simeq 1$ since we assume $M_H$ lies well 
below the $t\bar{t}$ threshold.  Here we use the framework and the notation of 
Refs.~\cite{KMR1,KMR2}\footnote{To be specific the signal-to-background ratio is given 
by comparing eqs.~(8,10) of \cite{KMR1} with eqs.~(8,17,19) of \cite{KMR2}.}.

Immediately we see a problem.  The small $p_T$ divergence of the dijet cross section, 
(\ref{eq:a3}), means that the background will be huge if just a missing-mass measurement on 
its own is performed.  We thus also require a central detector to impose a jet $p_T ({\rm 
or}~E_T)$ 
cut.  With such an additional detector we may select events with jets, say, with $E_T^2 > 
(3/16) M_H^2$.  In other words we trigger on double-diffractive events containing a pair of 
jets with angles $\theta > 60^\circ$ from the beam direction in the Higgs rest frame.  For a 
scalar Higgs boson this cut kills one half of the events, whereas the dijet cross section 
(\ref{eq:a3}) is reduced to
\be
\label{eq:a5}
\frac{d \hat{\sigma}}{dM^2} \; = \; \frac{9 \alpha_S^2}{8} \: \int_{3M^2/16}^{M^2/4} \: 
\frac{dp_T^2}{p_T^6} \: \left ( \frac{M^4}{4p_T^4} \: - \: \frac{M^2}{p_T^2} \right )^{- 
\frac{1}{2}} \; = \; 9.73 \: \frac{9 \alpha_S^2}{8 M^4}.
\ee
With the same scale in the couplings $\alpha_S$ in (\ref{eq:a4}) and (\ref{eq:a5}), and 
neglecting the NLO corrections\footnote{The NLO correction is not yet known for double-
diffractive dijet production from a colour-singlet state.  The corresponding $K$-factor is 
expected to be about the same, or a little larger, than that for Higgs production.}, we obtain 
the signal-to-background ratio
\bea
\label{eq:a6}
\frac{S}{B_{gg}} & = & \frac{\sqrt{2} G_F}{9.73 (81 \pi^2)} \: \frac{M^3}{\Delta M} \: 
\frac{1}{2} \: {\rm Br} (H \rightarrow b\bar{b}) \nonumber \\
& & \\
& \simeq & (4.3 \times 10^{-3}) \: {\rm Br} (H \rightarrow b\bar{b}) \: \left 
(\frac{M}{100~{\rm GeV}} \right )^3 \: \left ( \frac{250~{\rm MeV}}{\Delta M} \right ). 
\nonumber
\eea
The factor $\frac{1}{2} {\rm Br} (H \rightarrow b\bar{b})$ accounts for the branching ratio 
of the $H \rightarrow b\bar{b}$ decay and the $\theta > 60^\circ$ cut of the low $p_T$ jets.  
(The $H \rightarrow b\bar{b}$ branching ratio is about 0.7 if $M_H = 120~{\rm GeV}$.)

The ratio $S/B_{gg} \sim 5 \times 10^{-3}$ appears too small for the above approach to 
provide a viable signal for the Higgs boson.  However the situation is greatly improved if we 
are able to identify the $b$ and $\bar{b}$ jets.  If we assume that there is only a 1\% chance 
to misidentify a gluon jet as a $b$ jet, then tagging {\it both} $b$ and $\bar{b}$ will suppress 
the gluon background by $10^4$.  In this case only the true $b\bar{b}$ background will 
pose a problem.

\subsection{Signal-to-background ratio for $b$ quark dijets}

For the double-diffractive central production of a $b\bar{b}$ pair, the $H \rightarrow 
b\bar{b}$ signal/$b\bar{b}$ background ratio is much larger than that of Section~3.1.  Here 
the ratio is strongly enhanced due to the colour factors, gluon polarisation selection and the 
spin $\frac{1}{2}$ nature of the quarks.  First, the cross sections for inclusive colour-singlet 
dijet production are \cite{KMR2}
\bea
\label{eq:a7}
\frac{d \hat{\sigma}}{d\hat{t}} \: (gg \rightarrow gg) & = & \frac{9 \pi \alpha_S^2}{2 
p_T^4} \: \left ( 1 \: - \: \frac{p_T^2}{M^2} \right )^2 \nonumber \\
& & \\
\frac{d\hat{\sigma}}{d\hat{t}} \: (gg \rightarrow b\bar{b}) & \simeq & \frac{\pi 
\alpha_S^2}{6 p_T^2 M^2} \: \left (1 \: - \: \frac{2 p_T^2}{M^2} \right ). \nonumber
\eea
Since $p_T^2 < M^2/4$ the $b\bar{b}$ background is suppressed relative to the $gg$ 
background by
\be
\label{eq:a8}
\frac{d \hat{\sigma} (gg \rightarrow b\bar{b})}{d \hat{\sigma} (gg \rightarrow gg)} \; < \; 
\frac{1}{4 \times 27} \; < \; 10^{-2}.
\ee

Moreover, we emphasize that for the exclusive process the initial $gg$ state obeys special 
selection rules.  Besides being a colour-singlet, for forward outgoing protons the projection of 
the total angular momentum is $J_z = 0$ along the beam axis\footnote{This statement remains valid 
even when the blobs in Fig.~1, which describe the radiation of two $t$-channel gluons by the protons, 
include leading order BFKL or DGLAP ladders.}.  On the other hand, the Born 
amplitude for light fermion pair production\footnote{For light quark pair exclusive 
production, $p + p \rightarrow p + q\bar{q} + p$, with forward outgoing protons, the 
cancellation was first observed by Pumplin \cite{LUM20}, see also \cite{BC,KMR2}.} 
vanishes in this $J_z = 0$ state, see, for example, \cite{LUM23}.  This result follows from 
$P$- and $T$-invariance and fermion helicity conservation of the $J_z = 0$ amplitude 
\cite{LUM24}.  Thus, if we were to neglect the $b$-quark mass $m_b$, then at leading order 
we would have no QCD $b\bar{b}$-dijet background at all.  Even beyond LO, the 
interference between the signal and background amplitudes is negligibly small, since they 
have different helicity structure.  Therefore the form of the peak, observed in double-
diffractive exclusive $H \rightarrow b\bar{b}$ production, will not be affected by 
interference with $b\bar{b}$ jets produced by the pure QCD background process.

Of course, a non-vanishing $b\bar{b}$ rate is predicted when we allow for the quark mass or 
if we emit an extra gluon.  Nevertheless in the former case we still have an additional 
suppression to (\ref{eq:a8}) of about a factor of $m_b^2/p_T^2 \simeq 4 m_b^2/M_H^2 < 
10^{-2}$, whereas in the latter case the extra suppression is about $\alpha_S/\pi \simeq 0.05$.  
Note that events containing the third (gluon) jet may be experimentally separated from Higgs 
decay, where the two jets are dominantly co-planar\footnote{The situation here is similar to 
the signal-to-background ratio for intermediate mass Higgs production in polarised 
$\gamma\gamma$ collisions, which was studied in detail in \cite{LUM24,MSK}.}.  

Up to now we have discussed forward outgoing protons in the idealized case where their 
transverse momenta $\mbox{\boldmath $q$}_{1T}, \mbox{\boldmath $q$}_{2T} 
\rightarrow 0$.  In reality there is a strong correlation between these transverse momenta.  In 
particular, it implies that factorization (whereby the cross section is a product of Pomeron 
emission factors multiplied by the Pomeron-Pomeron fusion subprocess) is not valid 
\cite{LUM20}.  It is thus remarkable that the suppression of the double-diffractive 
$b\bar{b}$ dijet production (the QCD background process) is still valid for non-zero 
$q_{1T}$ and $q_{2T}$.  Indeed, the polarization vectors $\varepsilon_j$ of the \lq hard\rq\ 
gluons in Fig.~1 are directed along $(Q + q_i)_j$, where $j = 1,2$ denotes the two transverse 
components of the momenta.  In the case where we neglected $\mbox{\boldmath $q$}_{iT}$ 
and averaged over the direction of $\mbox{\boldmath $Q$}_T$ we obtained the polarization 
tensor
\be
\label{eq:b8}
\varepsilon_j \varepsilon_k \; \sim \; Q_j Q_k \; \sim \; \delta_{jk}^{(2)} \: \frac{Q_T^2}{2},
\ee
which corresponds to the $J_z = 0$ di-gluon state.  However, for non-zero $\mbox{\boldmath 
$q$}_{it}$ the tensor becomes
\be
\label{eq:c8}
\varepsilon_j \varepsilon_k \; \sim \; (Q + q_1)_j \: (Q + q_2)_k \; \sim \; \delta_{jk}^{(2)} \: 
\frac{Q_T^2}{2} \: + \: q_{1j} \: q_{2k}.
\ee
The linear term in $Q_T$ vanishes after the $\mbox{\boldmath $Q$}_T$ angular integration.  
In this way we obtain an admixture of $J_z = \pm 2$ di-gluon states, which leads to a 
contribution of the order of $4 q_{1T}^2 q_{2T}^2/Q_T^4$ to the $gg \rightarrow b\bar{b}$ 
cross section.  Thus the integral over the $Q_T$ loop in the amplitude (that is the dijet 
counterpart to (\ref{eq:b2})) becomes less infrared safe.  However the Sudakov-like form 
factor (\ref{eq:d2}) and the effective anomalous dimension, $\gamma > 0$, of the gluon $(xg (x, 
Q_T^2) \sim (Q_T^2)^\gamma)$ still suppress the contribution from the infrared domain 
\cite{KMR1}.  If we use the MRS gluon \cite{MRS}, we find the integral typically samples 
$Q_T$ values in the region $Q_T^2 \simeq 1.5 (3)~{\rm GeV}^2$ at the Tevatron (LHC) 
energy.  Thus the $|J_z| = 2$ admixture does not contribute more than 5\% (1.5\%) of the dijet 
cross section; here we take a mean $q_T$ of 400~MeV.  Of course the $|J_z| = 2$ 
contribution may be suppressed by selecting forward protons with smaller $q_T$.

So finally we see that identifying the $H \rightarrow b\bar{b}$ signal allows the background 
to be suppressed by more than a factor 3000.  The signal is thus in excess of the background, 
even for a mass resolution of $\Delta M \sim 4~{\rm GeV}$.  There is therefore an 
opportunity to see a clear peak at $M = M_H$, {\it provided} the cross section for 
double-diffractive Higgs production is large enough.

\subsection{The cross section for $p\porpbar \rightarrow p + H + \porpbar$}

The cross section for double-diffractive Higgs production at Tevatron and LHC energies has 
been calculated by several authors\footnote{A more complete set of references to related 
theoretical papers can be found in Ref.~\cite{AR}.} \cite{HIGGS,KMR1,EML,KMR3}.  If 
our recent perturbative QCD determination \cite{KMR3} is updated to incorporate the latest 
rapidity gap survival probability estimates\footnote{The gap survival probability for the 
double-diffractive processes is estimated to be $S^2 = 0.05$ at $\sqrt{s} = 2~{\rm TeV}$ and 
$S^2 = 0.02$ at $\sqrt{s} = 14~{\rm TeV}$.} of Ref.~\cite{SOFT}, then
\be
\label{eq:a9}
\sigma_H \; \equiv \; \sigma (p\bar{p} \rightarrow p + H + \bar{p}) \; \simeq \; 0.06 \: {\rm 
fb} \quad {\rm at} \quad \sqrt{s} \; = \; 2~{\rm TeV}
\ee
for a Higgs boson of mass 120~GeV.  Note that the value of cross section (\ref{eq:a9}) is 
comparable to the cross section generated by the $\gamma\gamma \rightarrow H$ fusion 
subprocess.  Recall that this QED contribution comes from large impact parameters, where 
the corresponding gap survival probability $S^2 = 1$ \cite{KMR3}.  $\sigma 
(\gamma\gamma \rightarrow H) $ is estimated to be about 0.03~fb at $\sqrt{s} = 2~{\rm 
TeV}$ and 0.3~fb at $\sqrt{s} = 14~{\rm TeV}$.  Note that the strong and electromagnetic 
contributions have negligible interference, because they occur at quite different values of the 
impact parameter:
\bea
\label{eq:b9}
\langle \rho_T^2 \rangle_{\rm em} & \gg & B_{\rm el} \; \sim \; 20~{\rm GeV}^{-2}, \nonumber \\
& & \\
\langle \rho_T^2 \rangle_{\rm str} & \sim & 4~{\rm GeV}^{-2}, \nonumber
\eea
where $B_{\rm el}$ is the $t$-slope of the elastic $pp$ cross section.

Prediction (\ref{eq:a9}) is lower than estimates made by other authors.  However it may be 
checked experimentally, since exactly the same mechanism, and calculation, is relevant to 
double-diffractive dijet production.  Indeed, double-diffractive dijet production, for jets with 
$E_T > 7~{\rm GeV}$, has been studied by CDF collaboration \cite{CDF}.  They find an 
upper limit for the cross section, $\sigma ({\rm dijet}) < 3.7~{\rm nb}$, as compared to our 
prediction of about $1~{\rm nb}$ \cite{KMR4}.  Using the dijet process as a monitor thus 
rules out the much larger predictions for $\sigma (p\bar{p} \rightarrow p + H + \bar{p})$ 
which exist in the literature.  Unfortunately the prediction $\sigma_H \simeq 0.06~{\rm fb}$ 
of (\ref{eq:a9}) means that Run II of the Tevatron, with an integrated luminosity of ${\cal L} 
= 15 {\rm fb}^{-1}$, should yield less than an event.  We should add that the 
double-diffractive Higgs search can also be made in the $\tau^+ \tau^-$ and $WW^*$ decay 
channels \cite{AR}, but, due to the small branching ratios, then the event rate is even less. 

On the other hand, the cross section $\sigma_H$, calculated in the perturbative QCD 
approach \cite{KMR1,KMR3,SOFT}, grows with energy and at $\sqrt{s} = 14~{\rm TeV}$ reaches 
$\sigma_H \simeq 2.2 {\rm fb}$ (corresponding to $d \sigma_H/dy \simeq 0.6~{\rm fb}$ at 
$y = 0$).  In fact, if we were to ignore the rapidity gap survival probability, $S^2$, then 
$\sigma_H$ would have increased by more than a factor of 100 in going from $\sqrt{s} = 
2~{\rm TeV}$ to $\sqrt{s} = 14~{\rm TeV}$.  However at larger energies the probability to 
produce secondaries which populate the gap increases and so $\sigma (pp \rightarrow p + H + 
p)$ increases only by a factor of 40.  Nevertheless there is a real chance to observe 
double-diffractive Higgs production at the LHC\footnote{Note that Refs.~\cite{KMR1,KMR3,SOFT} do 
not address the complications caused by multiple (or \lq pile-up\rq) interactions at the high 
luminosities at the LHC.}, since both the cross section and the 
luminosity are much larger than at the Tevatron.

\section{Double-diffractive $\chi$ meson production}

Of course, the missing-mass method may be used, not only for Higgs searches, but for many 
other double-diffractive exclusive reactions.  Particularly relevant examples are the 
production of scalar $(0^{++})$ $\chi_c$ and $\chi_b$ mesons 
\cite{AR,JP}.  These processes will allow another check, albeit qualitative, of our perturbative 
QCD techniques\footnote{We also have to include a suppression factor which 
represents the survival probability of the rapidity gaps.  This probability $S^2$ has been 
calculated for a range of diffractive processes in, for example, Ref.~\cite{SOFT}.} for 
calculating double-diffractive processes \cite{KMR3}.  In addition there are two reasons why 
$\chi$ production processes are of interest in their own right.  First, the production of $\chi$ 
mesons with a rapidity gap on either side ensures the selection of pure colour-singlet states, so 
there can be no admixture from a colour-octet production mechanism.  This should illuminate 
the dynamics of the hadroproduction of mesons containing heavy quarks.  Second, the data on 
inclusive $b\bar{b}$ production at the Tevatron lies about a factor of 3 above the 
conventional NLO QCD prediction.  A measurement of $\chi_b$ diffractive production is 
thus of interest.

As mentioned above, the cross section for double-diffractive $\chi$ meson production may be 
calculated by the same perturbative QCD approach that was applied to Higgs and dijet 
production \cite{KMR3}.  The $gg \rightarrow \chi$ vertex is given in terms of the width of 
the $\chi$ meson.  
We assume $\Gamma (\chi \rightarrow gg) 
\simeq \Gamma_{\rm tot} (\chi)$, with an observed width of $\Gamma_{\rm tot} = 14.9$~MeV 
\cite{PDG} for the $\chi_c (0^{++})$ meson.  
For $\chi_b (0^{++})$ we used the QCD lattice result \cite{CHIB} 
for the leading order $\Gamma (\chi \rightarrow gg)$ width of 354~keV.  
To be more precise, we included the standard NLO correction $(1 + 9.8 
\alpha_S/\pi)$ for $\chi_b (0^{++})$, see, for example, \cite{NLO}.  
We assume that the amplitude for $\chi$ production behaves as
\be
\label{eq:a10}
{\cal M} \; \propto \; e^{b (t_1 + t_2)/2}
\ee
with slope $b = 4~{\rm GeV}^{-2}$, where $t_i$ are the momentum transfer squared at the 
proton (antiproton) vertices.  For $\sqrt{s} = 2~{\rm TeV}$ we take the rapidity gap survival 
probability to be $S^2 = 0.05$ \cite{SOFT}.  With these input values, the double-diffractive 
cross sections estimated for Run II of the Tevatron are
\bea
\label{eq:a11}
\sigma \left (\chi_c (0^{++}) \right ) & \simeq & 600~{\rm nb}, \nonumber \\
& & \\
\sigma \left (\chi_b (0^{++}) \right ) & \simeq & 110~{\rm pb}. \nonumber
\eea
The corresponding rapidity distributions $d \sigma/dy$ are shown in Fig.~2.  

Note that the 
double-diffractive production of exclusive axial vector ($1^{++}$) and tensor $(2^{++})$ 
quarkonium states are strongly suppressed\footnote{The vanishing of the forward 
double-diffractive $1^{++}$ and $2^{++}$ quarkonium production was recently pointed out by 
F.\ Yuan \cite{FY} who used a non-relativistic formula for the evaluation of $P$-wave 
quarkonium decays.}.  The former results from the Landau-Yang theorem \cite{LY} which forbids 
the $1^{++} \rightarrow 2g$ transition for massless gluons.  The latter is directly related to 
the helicity-zero selection rule which we have discussed above.  It was known for a long time 
that in the non-relativistic limit the $J_z = 0$ amplitude for the $\gamma\gamma$ decay of the 
$2^{++}$ $^3 P_2$ positronium state vanishes \cite{AIA}.  This, of course, remains valid for the 
helicity-zero transition of a heavy tensor quarkonium state into two gluons \cite{KK}.

In reality, exclusive tensor $\chi$-meson production will occur due to corrections caused 
by relativistic effects (which are expected to be numerically small \cite{BHS}), as well as 
by the off-mass-shell corrections to the helicity-zero transition and by the admixture of 
$|J_z| = 2$ di-gluon states, which we explained in Section~3.2.  The largest 
contribution is expected from non-forward corrections, arising from the second term 
on the right-hand-side of (\ref{eq:c8}). The $\chi (2^{++})$ rate 
may therefore be as large as the fraction 
$0.2 \Gamma (2^{++})/\Gamma (0^{++})$ of the $\chi (0^{++})$ production cross sections 
of (\ref{eq:a11}).  As a consequence we anticipate a decrease of the $\chi (2^{++})$ 
cross section in the very forward region.  Note that the $J_z = 0$ 
selection rule becomes redundant for {\it inclusive} double-diffractive processes, so, 
for example, $2^{++}$ $\chi$-production will become more significant.

Of course the estimates of double-diffractive $\chi$ production are much less infrared 
stable than those for Higgs production.  For example, using GRV partons \cite{GRV} down 
to $Q^2 = 0.36~{\rm GeV}^2$, rather than MRS partons \cite{MRS} with frozen anomalous 
dimension for $Q^2 < 1~{\rm GeV}^2$, to calculate the gluon loop contribution at low 
virtuality\footnote{The contribution from the $Q^2 > 1~{\rm GeV}^2$ is approximately 
independent of whether GRV or MRS partons are used.}, enlarges the cross section by about 
a factor of 3 for $\chi_b$, and even larger for $\chi_c$.  It is known that the GRV gluon is 
rather too large in this domain.  However the comparison does demonstrate the infrared 
sensitivity of the estimates, which for such low mass particles should only be considered as an 
indication of the size of the cross section.  Nevertheless the $\chi$ cross sections are huge.  
For an integrated luminosity of ${\cal L} = 15~{\rm fb}^{-1}$ in Run II of the Tevatron we 
expect about $10^6~\chi_b$ events with a large rapidity gap on either side of the meson.

\section{Gluon factory}

The high event rate and the remarkable purity of the di-gluon system, that is generated in the 
exclusive double-diffractive production process, provides a unique\footnote{Rather we 
should say it is almost unique, since sometime ago it was proposed \cite{VAK} that a 
polarized $\gamma\gamma$ facility at a photon linear collider could also allow an 
experimental study of gluon jets via the process $\gamma\gamma \rightarrow gg$ in the $J_z 
= 0$ initial state.} environment to make detailed studies of high energy gluon jets.  Indeed we 
may speak of a \lq gluon factory\rq, since, as discussed in Section 3.2, double-diffractive 
quark di-jet production is strongly suppressed (by the $J_z = 0$ selection rule).  Recall that a 
precise comparison of quark and gluon jets requires that the two isolated gluon jets are 
produced from a point-like colour-singlet state\footnote{It is often said that a detailed study 
of the properties of gluon jets can be made by separating out the contribution of the $gg 
\rightarrow gg$ subprocess to a hadronic reaction.  In many cases this approach could be 
misleading since the $gg$ events do not originate from a pure colour-singlet system.  The 
interference between emissions from incoming and outgoing gluons leads to coherence 
effects which can strongly affect the final hadronic system, see, for example, 
\cite{KO,DKT}.} --- the counterpart of the celebrated $e^+ e^- \rightarrow q\bar{q}$ process 
--- see \cite{KO,KOW} for recent reviews.

Note that for QCD studies of gluon jets, the requirement of high resolution on the dijet mass 
is not essential.  Based on the estimates performed in Ref.~\cite{KMR3}, we expect, for 
example, about $10^5~gg$ events per day with $45 < E_T ({\rm jet}) < 55~{\rm 
GeV}$, at the LHC, assuming a luminosity of $1~{\rm fb}^{-1}$/day and neglecting 
corrections for acceptance and efficiency.  This number should 
be compared with the present experimental studies of the so-called unbiased gluon jets 
performed by the OPAL collaboration at LEP1 \cite{OPAL}.  Here gluon jets are identified in 
double-tagged $Z \rightarrow b\bar{b}g$ events.  Over about five years of running only 439 
pure gluon events were identified.

\section{Problems with a high-resolution pure missing-mass \\ method}

At first sight it looks attractive to use the missing-mass approach to search for (non-Standard 
Model) Higgs bosons which decay into invisible modes, such as gravitinos, neutralinos or 
gravitons.  The possibility of observing such an invisible Higgs in inclusive processes at the 
LHC was discussed recently in Ref.~\cite{EZ}.  Indeed in the exclusive channel, $pp 
\rightarrow p + H + p$, one should observe just two outgoing protons and nothing else but a 
sharp peak in the missing-mass distribution.  Unfortunately it is not clear how accurately it is 
possible to eliminate the background caused, for example, by QED radiation from the 
protons, or maybe even from radiative decays of low mass proton excitations.  The problem is 
that the probability to lose some energy by radiating photons of energy $\omega$ in the interval 
$d \omega$ is of order $(2 \alpha/3 \pi) (\langle q_T^2 \rangle/m^2) (d\omega/\omega)$ \cite{AB}, 
where $q_T$ is the transverse momentum of the outgoing proton of mass $m$.  Therefore the cross 
section for $pp \rightarrow (p \gamma) + (p \gamma)$, which may mimic a missing-mass event, is about 
\be
\label{eq:b11}
\left . \frac{d \sigma}{dy} \right |_{y = 0} \; \sim \; \left ( \frac{2 \alpha}{3 \pi} \: \frac{\langle 
q_T^2 \rangle}{m^2} \right )^2 \: \left ( \frac{dM^2}{M^2} \right ) \: \sigma_{\rm el} \; \sim \; 0.8~{\rm pb},
\ee 
where we have assumed $M = 120~{\rm GeV}, \Delta M = 250~{\rm MeV}$, and taken $\sigma_{\rm el} \simeq 
25~{\rm mb}$ and $\langle q_T^2 \rangle = 1/B_{\rm el} \simeq 0.05~{\rm GeV}^2$ at LHC energies.  This 
is huge in comparison to the cross section $d \sigma/dy_H \simeq 0.6~{\rm fb}$ for $pp \rightarrow p + 
H + p$.  Of course some of the QED radiation may be detected by an additional dedicated forward 
electromagnetic calorimeter\footnote{We thank A. Rostovtsev for discussions about photon detection.}.  
Also it is necessary to allow for the radiative tail which will spread out the shape of the Higgs peak.

This illustrates a basic problem in searching for {\it rare} processes using 
{\it high-resolution} missing-mass measurements which observe {\it only} the forward 
protons.  There will always be the possibility that part of the proton energy will escape 
undetected down, or near, the beam pipe.  Of course this deficiency would not exist if we 
were able to measure the products of the rare process with sufficient accuracy to confirm that 
their invariant mass coincides with the missing-mass measurement.  In this case the main 
purpose of the forward proton missing-mass detector would be to considerably improve 
the $\Delta M$ resolution.  For the gluon factory there is no problem, since high resolution is 
not essential and the event rate is enormous.  The missing-mass detector is used just to select 
large rapidity gap events.  Similarly there is no problem for double-diffractive $\chi$ meson 
production if the various decay modes are observed.

\section{Conclusion}

We have studied the proposal of triggering on forward going protons and antiprotons to 
perform a high resolution missing-mass search for the Higgs boson at the Tevatron, that is the 
process $p\bar{p} \rightarrow p + H + \bar{p}$ where a \lq plus\rq\ denotes a large rapidity 
gap.  However, we find that there is a huge QCD 
background arising from double-diffractive dijet production.  A central detector to trigger on 
large $E_T$ jets is essential.  Even so, the signal-to-background ratio is too small for a viable 
\lq missing-mass\rq\ Higgs search.  The situation is much improved if we identify $b$ and 
$\bar{b}$ jets.  The $gg \rightarrow H \rightarrow b\bar{b}$ signal is now in excess of the 
QCD $gg \rightarrow b\bar{b}$ background, even for a mass resolution of $\Delta M \sim 
4~{\rm GeV}$.  The only problem is that, when proper account is taken of the survival 
probability of the rapidity gaps, the $p\bar{p} \rightarrow p + H + \bar{p}$ event rate is too 
small at the Tevatron.  Recall that the experimental limit on the cross section for 
double-diffractive dijet production confirms the predicted small rates.

The pessimistic expectations of the missing-mass Higgs search at the Tevatron are, however, 
compensated by an interesting by-product of the double-diffractive proposal.  The 
double-diffractive production of dijets offers a unique {\it gluon factory}, generating huge 
numbers of essentially pure gluon jets from a colour-singlet state in an exceptionally clean 
environment.  Recall that the exclusive production of $q\bar{q}$ dijets, with a large rapidity 
gap on either side, is strongly suppressed by the $J_z = 0$ selection rule.

Finally, we give estimates for the double-diffractive production of $\chi_b$ and $\chi_c$ 
mesons at the Tevatron.  These mesons have smaller mass than the Higgs boson or dijet 
systems that we have considered, and so the QCD cross section estimates are much more 
qualitative.  Nevertheless the rates are huge, and the observation of $\chi_b$ and $\chi_c$ 
production (from colour-singlet states) should illuminate intriguing features of heavy flavour 
dynamics.

\section*{Acknowledgements}

We thank M. Albrow, J. Ellis, V. Fadin, R. Orava, A. Rostovtsev and W.J. Stirling for useful 
discussions.  VAK thanks the Leverhulme Trust for a Fellowship.  This work was also 
supported by PPARC, the Russian Fund for Fundamental Research (98-02-17629) and the EU 
Framework TMR programme, contract FMRX-CT98-0194 (DG-12-MIHT).

\newpage

\begin{figure} 
\begin{center} 
\psfig{figure=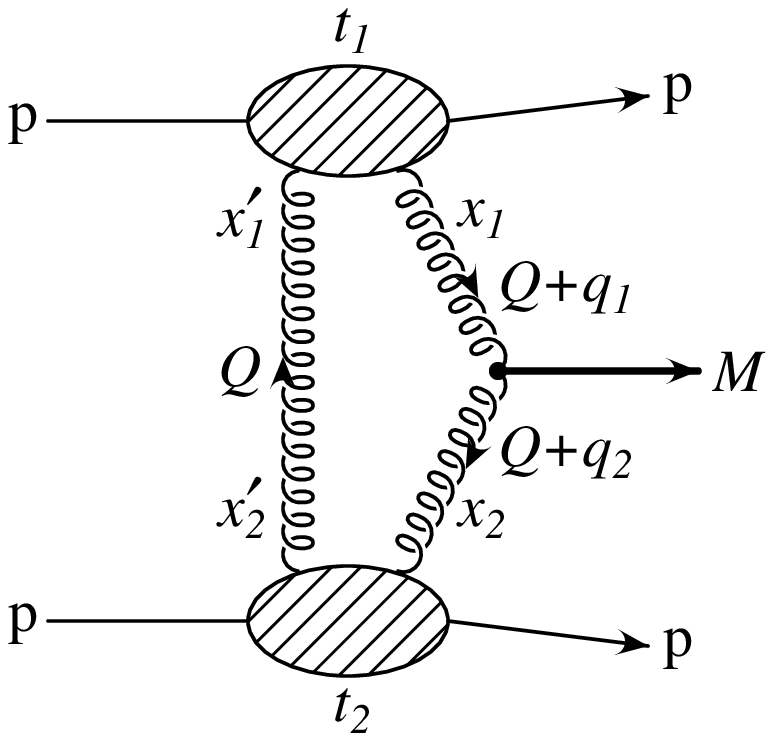,width=10cm} 
\end{center} 
\caption{Schematic diagram of double-diffractive production of a system 
of invariant mass $M$, that is the process $pp \rightarrow p + M + p$. 
\label{fig:Fig1} 
} 
\end{figure}

\begin{figure} 
\begin{center} 
\psfig{figure=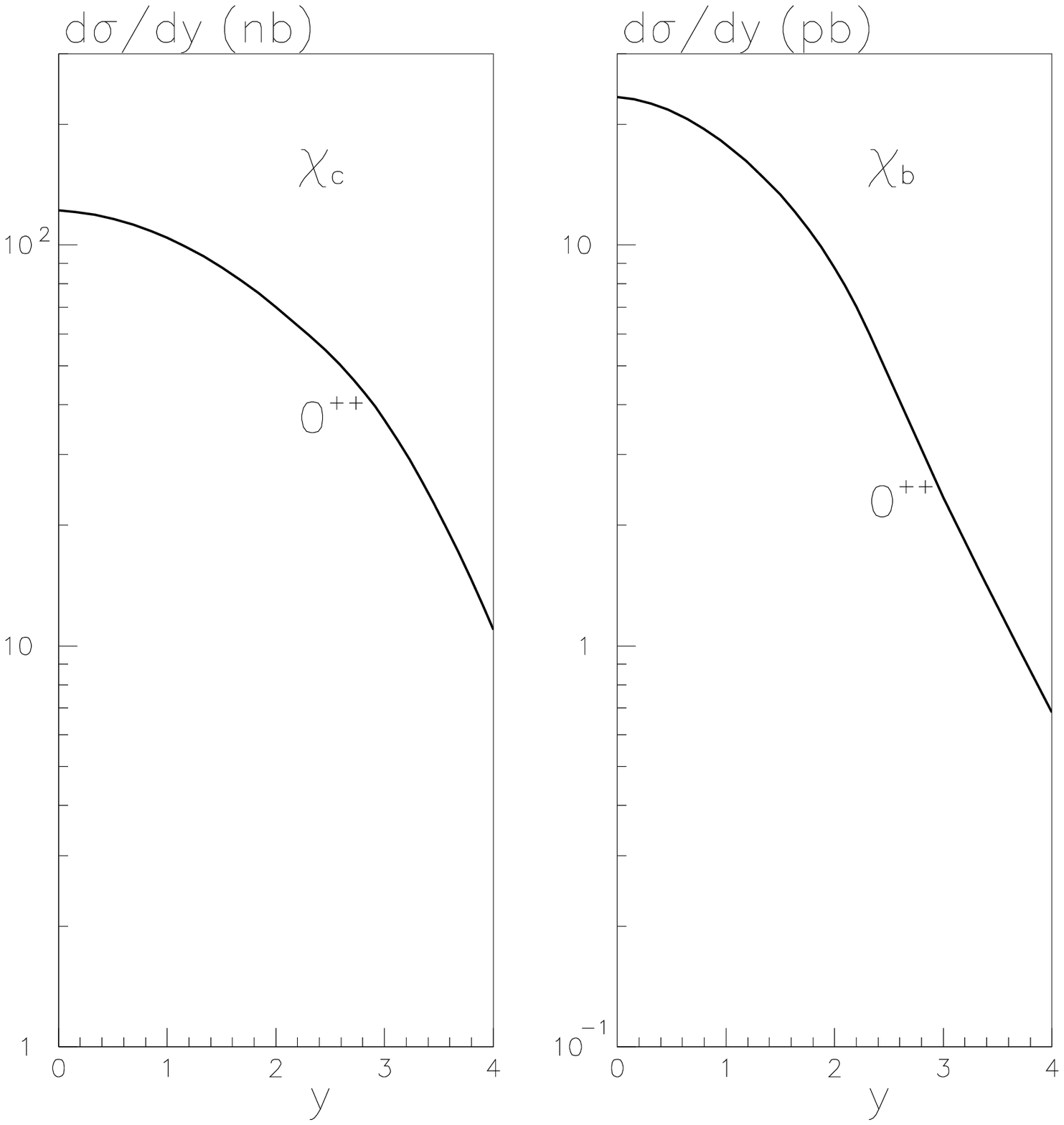,width=12cm} 
\end{center} 
\caption{The rapidity distributions for double diffractive $\chi_c$ 
and $\chi_b$ $(0^{++})$ production at the Tevatron.
\label{fig:Fig2} 
} 
\end{figure}


\newpage      
\begin{thebibliography}{xx}      
\bibitem{JE} J. Ellis, Lecture at 1995 Int. School of Subnuclear Physics, Erice (unpublished).
\bibitem{AR} M.G. Albrow and A. Rostovtsev, {\tt hep-ph/0009336}.
\bibitem{KMR2} A.D. Martin, M.G. Ryskin and V.A. Khoze, Phys. Rev. {\bf D56} (1997) 
5867.
\bibitem{HIGGS} A. Bialas and P.V. Landshoff, Phys. Lett. {\bf B256} (1991) 540.
\bibitem{KMR1} V.A. Khoze, A.D. Martin and M.G. Ryskin, Phys. Lett. {\bf B401} (1997) 
330.
\bibitem{EML} E.M. Levin, {\tt hep-ph/9912403}.
\bibitem{KL} D. Kharzeev and E.M. Levin, Phys. Rev. {\bf D63} (2001) 073004.
\bibitem{KMR3} V.A. Khoze, A.D. Martin and M.G. Ryskin, Eur. Phys. J. {\bf C14} (2000) 
525.
\bibitem{BE} A. Berera, Phys. Rev. {\bf D62} (2000) 014015.
\bibitem{BC} A. Berera and J.C. Collins, Nucl. Phys. {\bf B474} (1996) 183.
\bibitem{LUM20} J. Pumplin, Phys. Rev. {\bf D52} (1995) 1477.
\bibitem{LUM23} K.A. Ispiryan, I.A. Nagorskaya, A.G. Oganesyan and V.A. Khoze, Sov. J. 
Nucl. Phys. {\bf 11} (1970) 712.
\bibitem{LUM24} D.L. Borden, V.A. Khoze, W.J. Stirling and J. Ohnemus, Phys. Rev. {\bf 
D50} (1994) 4499; \\
V.S. Fadin, V.A. Khoze and A.D. Martin, Phys. Rev. {\bf D56} (1997) 484.
\bibitem{MSK} M. Melles, W.J. Stirling and V.A. Khoze, Phys. Rev. {\bf D61} (2000) 
054015.
\bibitem{MRS} A.D. Martin, R.G. Roberts and W.J. Stirling, Phys. Lett. {\bf B387} (1996) 
419.
\bibitem{SOFT} V.A. Khoze, A.D. Martin and M.G. Ryskin, Eur. Phys. J. {\bf C18} (2000) 167.
\bibitem{CDF} CDF Collaboration:  T. Affolder et al., Phys. Rev. Lett. {\bf 85} (2000) 4215.
\bibitem{KMR4} V.A. Khoze, A.D. Martin and M.G. Ryskin, {\tt hep-ph/0006005}; 
{\tt hep-ph/0007083}, Phys. Lett. {\bf B502} (2001) 87.
\bibitem{JP} J. Pumplin, Phys. Rev. {\bf D47} (1993) 4820.
\bibitem{PDG} Review of Particle Properties, Eur. Phys. J. {\bf C15} (2000) 1.
\bibitem{CHIB} S. Kim, Nucl. Phys. Proc. Suppl. {\bf 47} (1996) 437.
\bibitem{NLO} R. Barbieri, M. Caffo, R. Gatto and E. Remiddi, Phys. Lett. {\bf B95} (1980) 
93; Nucl. Phys. {\bf B192} (1981) 61.
\bibitem{FY} F. Yuan, {\tt hep-ph/0103213}.
\bibitem{LY} L.D. Landau, Dokl. Akad. Nauk. SSSR {\bf 60} (1948) 213; \\
C.N. Yang, Phys. Rev. {\bf 77} (1950) 242.
\bibitem{AIA} A.I. Alekseev, Sov. Phys. JETP {\bf 34} (1958) 826.
\bibitem{KK} M. Krammer and H. Krasemann, Phys. Lett. {\bf B73} (1978) 58.
\bibitem{BHS} L. Bergstr\"{o}m, G. Hulth and H. Snellman, Z. Phys. {\bf C16} (1983) 263; \\
Z.P. Li, F.E. Close and T. Barnes, Phys. Rev. {\bf D43} (1991) 2161.
\bibitem{GRV} M. Gl\"{u}ck, E. Reya and A. Vogt, Eur. Phys. J. {\bf C5} (1998) 461.
\bibitem{VAK} V.A. Khoze, Plenary talk at Workshop on Physics with Linear Colliders, 
Saariselka, Finland, 9-14 Sept., 1991; published in Saariselka Workshop 1991, 547.
\bibitem{KO} V.A. Khoze and W. Ochs, Int. J. Mod. Phys. {\bf A12} (1997) 2949.
\bibitem{DKT} Yu.L. Dokshitzer, V.A. Khoze and S.I. Troyan, Sov. J. Nucl. Phys. {\bf 46} 
(1987) 712.
\bibitem{KOW} V.A. Khoze, W. Ochs and J. Wosiek, {\tt hep-ph/0009298}.
\bibitem{OPAL} OPAL Collaboration:  G. Abbiendi et al., Eur. Phys. J. {\bf C11} (1999) 
217.
\bibitem{EZ} O.J.P. \'{E}boli and D. Zeppenfeld, Phys. Lett. {\bf B495} (2000) 147.
\bibitem{AB} A.I. Akhiezer and V.B. Berestetskii, Quantum Electrodynamics (Interscience, New York, 1965).
\end{thebibliography}
\end{document}